\documentclass[twocolumn,showpacs,amsmath,amssymb,pra,superscriptaddress]{revtex4-1}
\usepackage{graphicx}
\usepackage{epstopdf}
\usepackage{epsfig}
\usepackage{dcolumn}
\usepackage{color}
\usepackage{bm}
\usepackage{mathrsfs}
\usepackage{bbold}
\usepackage{amssymb}
\usepackage{upgreek}
\DeclareMathOperator{\Tr}{Tr}

\begin{document}

\title{The short-time and long-time behaviors of non-Markovianity measure using
two-time correlation functions in open quantum systems}

\author{Md.~Manirul Ali}
\email{mani@mail.ncku.edu.tw}
\affiliation{Department of Physics, National Cheng Kung University, Tainan 70101, Taiwan}
\author{Ping-Yuan Lo}
\affiliation{Department of Physics, National Cheng Kung University, Tainan 70101, Taiwan}
\author{Matisse Wei-Yuan Tu}
\affiliation{Department of Physics, National Cheng Kung University, Tainan 70101, Taiwan}
\author{Wei-Min Zhang}
\email{wzhang@mail.ncku.edu.tw}
\affiliation{Department of Physics, National Cheng Kung University, Tainan 70101, Taiwan}

\date{\today}

\begin{abstract}
We investigate non-Markovianity measure using two-time correlation functions for open quantum systems.
We define non-Markovianity measure as the difference between the exact two-time correlation function
and the one obtained in the Markov limit. Such non-Markovianity measure can easily be measured in experiments.
We found that the non-Markovianity dynamics in different time scale crucially depends on the
system-environment coupling strength and other physical parameters such as the initial temperature
of the environment and the initial state of the system. In particular, we obtain the short-time and
long-time behaviors of non-Markovianity for different spectral densities. We also find that the thermal
fluctuation always reduce the non-Markovian memory effect. Also, the non-Markovianity measure shows
non-trivial initial state dependence in different time scales.
\end{abstract}

\maketitle

\section{Introduction}\label{sec:introduction}

Markov approximation \cite{Scully97,Carmichael99,Gardiner04,Milburn08} which completely ignores the
memory effects between the system and its environments is widely used in the study
of open quantum systems. However, in quantum information processing, many quantum
devices exhibit {\it non-Markovian} (memory) behaviors which prohibits the use of Markov approximation.
In particular, non-Markovian memory effects are typically significant for structured
environments at low temperature in the strong system-environment coupling regime. Various
concepts of non-Markovianity were introduced  recently to define the border between Markovian
and non-Markovian quantum evolution, although the very definition of non-Markovianity is still
a debated issue. Many different measures of {\it non-Markovianity} have been proposed in the literature to
quantify memory effects in open systems, based on, for examples, the divisibility of dynamical
map \cite{divisibility1,divisibility2,divisibility3,reviewPlenio}, the distinguishability of states \cite{distinguishability1,distinguishability2,distinguishability3,distinguishability4},
quantum entanglement \cite{entanglement1,entanglement2,entanglement3,entanglement4,entanglement5},
quantum Fisher information \cite{information1,information2,information3,information4}, mutual
information \cite{mutual1,mutual2,mutual3,mutual4}, geometrical characterization
\cite{reviewPlenio,geometry} and the decay rate of the master equation itself
\cite{decayrate1,decayrate2}. Predominately, almost all these measures of non-Markovianity
are introduced in terms of mathematical quantities. Most of them do not directly related to
physically measurable observables, although some of them are experimentally measured through
reconstruction of the output state in terms of density matrix, using quantum state tomography
\cite{distinguishability3,NatComn}. Also, some of these measures even quantitatively
diverge from each other \cite{mutual4,diverge,nvtyVacchini2014}.

Recently, a general theory of non-Markovian dynamics is developed for open quantum systems of bosons (fermions)
interacting with a general bosonic (fermionic) environments through particle-tunneling processes \cite{general},
where exact dynamics of open quantum systems is explored through the nonequilibrium Green's functions which account
all the information of non-Markovian back-action memory effects. The relation between the exact master equation
and the nonequilibrium Green's functions manifests the fundamental importance of correlation functions in the
description of the non-Markov dynamics for open systems. Therefore, non-Markovianity measure extracted from correlation
functions in time-domain has a special significance. In particular, two-time correlation functions are experimentally
measurable. For examples, the two-time correlation functions of the electromagnetic field emitted by an atom can be
measured through fluorescence spectrum \cite{Scully97,Carmichael99,Gardiner04,Milburn08}; the two-time correlation
functions of the number of emitted photons give the information about the photon statistics and describe the behavior
of photon bunching and antibunching \cite{Scully97,Carmichael99,Gardiner04,Milburn08}; two-time correlation functions
have also been measured in optical measurements and light-harvesting photosynthesis for testing the non-Markovian
memory effects \cite{malik,engel}; the two-time correlation functions of the electric current through nanostructure
devices are used in quantum transport to study the current fluctuations and noise spectrum
\cite{noiseS1,noiseS2,PeiYun}. Intuitively, two-time correlation functions
correlating a past event with its future provide useful information about the system-environment
back-action processes revealing the non-Markovian memory effects.

In this paper, we introduce a non-Markovianity measure using two-time correlation functions.
In open quantum systems, when the Markov approximation \cite{Scully97,Carmichael99,Gardiner04,Milburn08}
is valid, two-time correlation functions are usually calculated using quantum regression theorem
where the reservoirs are assumed to be always in an equilibrium state. Our non-Markovianity measure
in terms of two-time correlation functions is defined as the departure of the exact dynamics of
the two-time correlation function from the one obtained through quantum regression theorem in the
Markov limit. Our definition of non-Markovianity measure is valid for arbitrary finite temperature of
the environment. We also show that the non-Markovianity defined in this way is a function of time and
therefore the non-Markovianity measure is a dynamical quantity, which is physically expected. The non-Markovianity
defined in this method also depends crucially on various system and environment parameters and on the
nature of system-environment coupling. As a specific example, we take an open system with one dynamical
degree of freedom coupled to a general non-Markovian environment containing infinite degrees of freedom
described by the famous Fano-Anderson model \cite{FanoAnd1,FanoAnd2}. Then the two-time correlation
function we used to define the non-Markovianity measure is just the first-order coherence function in
quantum optics, which can be easily measured in experiments \cite{Scully97,Carmichael99,Gardiner04,Milburn08,noiseS1,noiseS2,PeiYun}.

The rest of the paper is organized as follows. In Sec.~\ref{subsec:exact}, we consider our open system described
by the Fano-Anderson model, where we show that the exact two-time correlation functions of system operators can be
obtained through exact master equation associated with the nonequilibrium Green's functions. In
Sec.~\ref{subsec:qrt}, we calculate the two-time correlation function in Markov limit using the quantum
regression theorem. We define our non-Markovianity measure in Sec.~\ref{subsec:measure} using two-time
correlation functions, where we also include a comparative discussion of our definition
with other relevant non-Markovianity measure. In Sec.~\ref{sec:results}, we present our numerical results
and discussions, the dynamics of our non-Markovianity measure for different system-environment coupling,
different initial temperature of the environment, and different initial states of the system. Our results also
show non-trivial interesting non-Markovian features on the dependence of the system and
environmental parameters, in particular, the existence of both the short-time and long-time
behavior of the non-Markovianity. Finally, a conclusion is given in Sec.~\ref{sec:conclusion}.

\section{The non-Markovianity measure based on two-time correlation function}\label{sec:main}

\subsection{Two-time correlation function obtained from the exact master equation}\label{subsec:exact}

We consider an open system described by the famous Fano-Anderson Hamiltonian \cite{FanoAnd1,FanoAnd2}
\begin{eqnarray}
\hskip -0.5cm
\label{Tot2}
H =  \hbar \omega_0 a^{\dagger} a + \sum_k \hbar \omega_k b_k^{\dagger} b_k
+ \sum_k \hbar V_k \left( a^{\dagger} b_k + a b_k^{\dagger}  \right),
\end{eqnarray}
where $a^{\dagger}$ and $a$ are the creation and annihilation operators of a single frequency mode.
A general non-Markovian environment is modeled as a collection of infinite modes, and $b_k^{\dagger}$
and $b_k$ are the corresponding creation and annihilation operators of the $k$-th mode with frequency
$\omega_k$. The parameter $V_k$ is the tunneling amplitude between the system and its environment.
The model Hamiltonian (\ref{Tot2}) has wide applications in atomic physics and condensed matter physics
\cite{FanoAnd3,FanoAnd4}.

We consider initially the environment be at thermal equilibrium. Tracing over all the environmental
degrees of freedom using the influence functional approach \cite{influence1,influence2} in the
coherent state representation \cite{coherent}, the exact master equation of such open systems has been
derived \cite{bosonic,annphys,njp14}:
\begin{align}
\label{master2}
\frac{d}{dt} \rho(t)=& - i \omega_0^{\prime} (t) \left[ a^{\dagger} a , \rho(t) \right] \notag \\
&{} + \gamma(t) \left[ 2 a \rho(t) a^{\dagger} - a^{\dagger} a \rho(t) - \rho(t) a^{\dagger} a \right] \notag \\
&{} + \widetilde{\gamma}(t) \left[ a \rho(t) a^{\dagger} + a^{\dagger} \rho(t) a - a^{\dagger} a \rho(t)
- \rho(t) a a^{\dagger} \right],
\end{align}
The time-dependent coefficients in the master equation are fully determined by the nonequilibrium Green's
functions $u(t,t_0)$ and $v(t,t)$, which satisfy the following Dyson equation of motion and the
non-equilibrium fluctuation-dissipation theorem \cite{general}, respectively,
\begin{eqnarray}
\label{ide}
{\dot u}(t,t_0) + i \omega_0 u(t,t_0) + \int_{t_0}^t d\tau g(t,\tau) u(\tau,t_0) = 0, \\
\hskip -0.2cm
v(t,t+\tau)\!=\!\!\!\int_{t_0}^t\!\!\!\!d\tau_1\!\!\int_{0}^{t+\tau}\!\!\!\!\!\!\!\!\!d\tau_2 u(t,\tau_1) {\tilde g}(\tau_1,\tau_2)u^{\ast}(t+\tau,\tau_2).
\label{vtb}
\end{eqnarray}
The integral kernels in Eqs.~(\ref{ide}) and (\ref{vtb}) characterize all the non-Markovian back-action
between the system and the reservoir, and can be determined by the spectral density $J(\omega)$ of the reservoir
through the relations: $g(t,\tau) = \int_0^{\infty} \frac{d\omega}{2\pi} J(\omega) e^{-i\omega(t-\tau)}$,
and ${\tilde g}(\tau_1,\tau_2) = \int_0^{\infty} \frac{d\omega}{2\pi} J(\omega) {\bar n}(\omega,T)
e^{-i\omega(\tau_1-\tau_2)}$, where the spectral density is defined by
$J(\omega)=2\pi\sum_k |V_k|^2 \delta(\omega-\omega_k)$ and ${\bar n}(\omega,T)=\frac{1}{e^{\hbar \omega / k_B T}-1}$
is the particle number distribution of the bosonic reservoir.

The functions $u(t,t_0)$ and $v(t,t+\tau)$ are indeed related to the two-time correlation functions
$\langle[a(t),a^{\dagger}(t_0)]\rangle$ and $\langle a^{\dagger}(t+\tau) a(t)\rangle$ respectively,
which are the two basic nonequilibrium Green's functions in the Schwinger-Keldysh theory \cite{general}.
These two-time correlation functions can also be calculated directly from equation of motion approach
\cite{eom1,eom2}. In fact, all two-time correlation functions of system operators can be constructed in terms of the above
nonequilibrium Green's functions. For example, as we have shown in our previous works \cite{annphys,njp10},
the exact two-time correlation function $\langle a^{\dagger}(t) a(t+\tau)\rangle$ can be expressed
in terms of $u(t,t_0)$ and $v(t,t+\tau)$ as follows,
\begin{eqnarray}
\nonumber
\langle a^{\dagger}(t) a(t+\tau) \rangle_{\cal E}\!=\!u^{\ast}(t,t_0) n(t_0) u(t+\tau,t_0)\!+\!v^{\ast}(t,t+\tau). \\
\label{exact2}
\end{eqnarray}
The exact analytic solution of the integro-differential equation (\ref{ide}) is recently given in \cite{general},
\begin{eqnarray}
\nonumber
u(t,t_0) = \mathcal{Z} e^{-i \omega_b (t-t_0)}\!+\!\!\!\int_0^\infty\!\!\!\!\!\!
\frac{J(\omega) e^{-i\omega(t-t_0)}}{\left[\omega - \omega_0 -
\Delta(\omega) \right]^2 + \gamma^2(\omega)}\frac{d\omega}{2\pi}, \\
\label{ut2}
\end{eqnarray}
where $\Delta(\omega) = {\cal P}\int_0^\infty \frac{J(\omega')}{\omega-\omega'} \frac{d\omega'}{2\pi}$ is
a principal-value integral and $\gamma(\omega)=J(\omega)/2$, which are the real and imaginary parts of
the self-energy correction to the system induced by the system-environment coupling,
\begin{eqnarray}
\hskip -0.5cm
\Sigma(\omega \pm i 0^{+})=\int_0^\infty \frac{d\omega^{\prime}}{2\pi}
\frac{J(\omega^{\prime})}{\omega-\omega^{\prime} \pm i 0^{+}}=\Delta(\omega) \mp i \gamma(\omega).
\label{sigmaz}
\end{eqnarray}
The first term in Eq.~(\ref{ut2}) is the contribution of the dissipationless localized
mode, which exists only when the environment has band-gap structures. The localized mode frequency
$\omega_b$ is determined by the pole condition $\omega_b - \omega_0 - \Delta(\omega_b)=0$, and
$\mathcal{Z}=\left[ 1 - \Sigma^{\prime}(\omega_b) \right]^{-1}$ corresponds to the residue
at the pole, which gives the amplitude of the localized mode. Thus, the two-time correlation
functions can be exactly calculated from the exact solution of Eq.~(\ref{ut2}) through Eq.~(\ref{vtb}).

\subsection{Two-time correlation functions in Markov limit using quantum regression theorem}\label{subsec:qrt}

Two-time correlation functions can easily be obtained through the {\it quantum regression theorem} defined
in the Markov limit \cite{Lax} when the back-action or memory effect is totally negligible.
Quantum regression theorem states that in the Markov limit, evolution equations of the two-time
correlation functions of system observables is the same as the evolution equation for single-time
expectation values of the observables \cite{Carmichael99,Lax}. The single time expectation values of a
complete set of system operators ${\hat O}_i$, can be calculated using the master equation (\ref{master2})
\begin{eqnarray}
&&\frac{\partial}{\partial t} \langle {\hat O}_i(t) \rangle = \Tr_S \left[ {\hat O}_i(0)
\frac{\partial}{\partial t} \rho(t) \right] = \sum_{j} M_{ij} \langle {\hat O}_j(t) \rangle,
\label{meanvalue1}
\end{eqnarray}
where the evolution matrix elements $M_{ij}$ are determined from the master equation (see Appendix-\ref{sec:qrtappendix}).

According to quantum regression theorem, the evolution equation (\ref{meanvalue1}) for single-time
expectation values is also valid for the two-time correlation functions in the Markov limit:
\begin{eqnarray}
\frac{\partial}{\partial \tau} \langle {\hat O}_1(t) {\hat O}_i (t+\tau) \rangle = \sum_{j} M_{ij}
\langle {\hat O}_1(t) {\hat O}_j (t+\tau)\rangle.
\label{theorem}
\end{eqnarray}
The operator ${\hat O}_1$ can be any system operator, not necessarily one of the $\{{\hat O}_i\}$, and
$M_{ij}$ is the same matrix given in Eq.~(\ref{meanvalue1}).

The Markov master equation for the open system of Eq.~(\ref{Tot2}) is well known in the literature
\cite{Carmichael99} and it has the same form as the exact master equation (\ref{master2}). However, the
coefficients in the Markov master equation become time-independent, and can be derived from the exact master
equation (\ref{master2}) using a perturbation expansion up to the second order and then taking the Markov
limit \cite{bosonic}. The result is,
\begin{eqnarray}
\gamma(t) &\rightarrow& \gamma = J(\omega_0)/2, \\
\widetilde{\gamma}(t) &\rightarrow& \widetilde{\gamma} = J(\omega_0){\bar n}(\omega_0,T), \\
\omega_0^{\prime}(t) &\rightarrow& \omega_0 + \delta \omega_0,
\end{eqnarray}
with $\delta \omega_0 = {\cal P}\int_{0}^{\infty} \frac{d\omega}{2\pi}\frac{J(\omega)}{(\omega-\omega_0)}$.
Using this Markov master equation, one can obtain the
evolution matrix $M_{ij}$ in the complete set of system operators ($a$, $a^{\dagger}$, $a^{\dagger}a$, $\mathbb{1}$):
\begin{equation}
{\bf M} =
\left( \begin{array}{cccc}
\beta & 0 & 0  & 0 \\
0 & \beta^{\ast} & 0 & 0 \\
0 & 0 & -2\gamma & {\tilde \gamma} \\
0 & 0 & 0 & 0
\end{array} \right),
\label{matrix2}
\end{equation}
where $\beta = -\left( \gamma + i \omega_0^{\prime} \right)$. By choosing the operator ${{\hat O}_1}=a^\dagger$,
the evolution equation for the two-time correlation function $\langle a^{\dagger}(t) a(t+\tau) \rangle$
based on quantum regression theorem is given by
\begin{eqnarray}
\nonumber
&&\frac{\partial}{\partial\tau} \langle a^{\dagger}(t) a(t+\tau) \rangle_{\cal M} \\
&&{} = -\left( \gamma + i \omega_0^{\prime} \right) \langle a^{\dagger}(t) a(t+\tau) \rangle_{\cal M},
\label{qrtm1}
\end{eqnarray}
which can be solved explicitly with the solution
\begin{eqnarray}
\nonumber
\hskip -0.6cm
&&\langle a^{\dagger}(t) a(t+\tau) \rangle_{\cal M} \\
&&{} \!=\! \left[ n(t_0) e^{-2\gamma t} + {\bar n}(\omega_0,T) \left( 1 - e^{-2\gamma t} \right) \right] e^{-(\gamma + i \omega_0^{\prime}) \tau}\!\!.
\label{qrtm2}
\end{eqnarray}
Here ``${\cal M}$'' means that the correlation function is calculated in the Markov limit through the
quantum regression theorem. In the next subsection, we will use the exact two-time correlation functions
and the one in the Markov limit to define the non-Markovianity measure.

\subsection{Definition of non-Markovianity and its measure}\label{subsec:measure}

Our definition of the non-Markovianity measure quantifies the departure of the exact dynamics of two-time correlation
function from the one obtained in the Markov limit. Thus, non-Markovianity measure in terms of two-time
correlation functions is defined by
\begin{eqnarray}
{\cal N}(t,\tau) = \left|  g_{\cal E}(t,\tau) - g_{\cal M}(t,\tau)  \right|,
\label{nonM}
\end{eqnarray}
where $g_{\cal E}(t,\tau)$ and $g_{\cal M}(t,\tau)$ are the first order coherence functions in terms of two-time
correlation functions for the {\it exact} and {\it Markov} dynamics, respectively,
\begin{eqnarray}
g_{\cal E}(t,\tau) = \frac{\langle a^{\dagger}(t) a(t+\tau) \rangle_{\cal E}}{\sqrt{\langle a^{\dagger}(t) a(t)
\rangle_{\cal E}~\langle a^{\dagger}(t+\tau) a(t+\tau) \rangle_{\cal E}}},
\end{eqnarray}
and
\begin{eqnarray}
g_{\cal M}(t,\tau) = \frac{\langle a^{\dagger}(t) a(t+\tau) \rangle_{\cal M}}{\sqrt{\langle a^{\dagger}(t) a(t)
\rangle_{\cal M}~\langle a^{\dagger}(t+\tau) a(t+\tau) \rangle_{\cal M}}},
\end{eqnarray}
Here $\langle a^{\dagger}(t) a(t) \rangle_{\cal E}$ and $\langle a^{\dagger}(t) a(t) \rangle_{\cal M}$
can be calculated simply from Eqs.~(\ref{exact2}) and (\ref{qrtm2}) by setting $\tau=0$. In other words,
we take $g_{\cal M}(t,\tau)$, which is obtained in the Markov limit, as a reference point to account the
non-Markovianity of the open system dynamics.

Based on this definition (\ref{nonM}), it is clear that such a non-Markovianity measure can be directly measured
in experiments through the measurement of the exact two-time correlation function or more precisely the
first order coherence function $g_{\cal E}(t,\tau)$. The two-time correlation functions are defined for
arbitrary finite temperature of the environment. Hence the non-Markovianity measure defined in Eq.~(\ref{nonM})
take into account the proper temperature dependence, which will be explicitly described in Sec.~\ref{sec:results}.
Our non-Markovianity measure is also an explicit function of time. Physically, this must be the case because
non-Markovianity characterizes the memory effect which must be time-dependent in general. It is also clear from
the analytical expression (\ref{nonM}) through Eq.~(\ref{exact2}) that this non-Markovianity measure
${\cal N}(t,\tau)$ depends on the initial state of the system. This initial state dependence
is indeed expected as a typical non-Markovian memory effect.

Here, we should point out that a similar definition of non-Markovianity measure has been recently proposed
by Guarnieri {\it et al.}~for a pure dephasing spin-boson model at zero temperature \cite{nvtyVacchini2014}.
In their work, the non-Markovianity was defined as
\begin{eqnarray}
{\cal N}^{\prime} = 1 - \frac{\langle A(t_2)B(t_1) \rangle_{qrt}}{\langle A(t_2) B(t_1) \rangle_{exc}},
\end{eqnarray}
where $\langle A(t_2)B(t_1) \rangle_{qrt}$ is the two-time correlation function obtained from quantum regression
theorem by naively using the exact master equation, and $\langle A(t_2) B(t_1) \rangle_{exc}$ is the exact
two-time correlation function. We find that this definition cannot account all the non-Markovianity. This is because
quantum regression theorem is based on a crucial assumption that the total density operator of the system plus
environment must be approximately factorized and the environment must always remain in the initial state at
an arbitrary time, namely
\begin{eqnarray}
\rho_{tot}(t)=\rho(t)\otimes\rho_E(t_0).
\label{factorized}
\end{eqnarray}
The detail derivation of this condition is given in Appendix-\ref{sec:qrtappendix}). For non-Markov evolution
based on the exact master equation, the condition of Eq.~(\ref{factorized}) cannot be satisfied in general
and therefore the two-time correlation functions calculated from such a naive quantum regression theorem is
inadequate. Thus, non-Markovianity defined in Ref.~\cite{nvtyVacchini2014} cannot account the true non-Markovianity
degree due to the improper use of the quantum regression theorem.

To show this inadequacy explicitly, let us apply this naive quantum regression theorem defined in
Ref.~\cite{nvtyVacchini2014} to our exact master equation (\ref{master2}). It is straightforward to show
(see Appendix-\ref{sec:qrtexac}) that the two-time correlation function based on such a quantum regression
theorem is given by
\begin{eqnarray}
\nonumber
\langle a^{\dagger}(t) a(t+\tau) \rangle_{qrt} &=& u^{\ast}(t,t_0) n(t_0) u(t,t_0) u(\tau,t_0) \\
&&{} +  v(t,t) u(\tau,t_0).
\label{qrte3}
\end{eqnarray}
This result is different from Eq.~(\ref{exact2}) which is exact, it is also different from Eq.~(\ref{qrtm2})
which is the two-time correlation function in the Markov limit. Thus, the two-time correlation function of
Eq.~(\ref{qrte3}), if it is applicable, contain partial non-Markovianity. As a result, the
non-Markovianity measure given in Ref.~\cite{nvtyVacchini2014} cannot account all the non-Markovianity degree.

On the other hand, in the literature, there are some works \cite{Goan1,Goan2,Dara} that attempted to generalize quantum
regression theorem to non-Markovian regime where the system and the reservoirs are only weakly coupled to each other
so that the master equation can be obtained perturbatively under the condition of Eq.~(\ref{factorized}). In this
case Eq.~(\ref{factorized}) is actually the well-known Born-approximation. Thus, under this condition, quantum regression
theorem is valid so that one can calculate two-time correlation functions from the Born-approximated master equation.
But, the result obtained from such a quantum regression theorem cannot be used as a reference point to define non-Markovianity measure, because such a quantum regression theorem-based two-time
correlation functions may contain partial non-Markovianity. In conclusion, our definition of non-Markovianity
properly accounts all the non-Markovianity degree in terms of a physically measurable quantity.

\section{Results and Discussion}\label{sec:results}

In order to evaluate explicitly the two-time correlation functions and the non-Markovianity measure ${\cal N}(t,\tau)$,
we need to specify the spectral density $J(\omega)$ of the environment. In the following, we consider the Ohmic-type
spectral density which can simulate a large class of thermal bath \cite{LeggettRMP}
\begin{eqnarray}
J(\omega) = 2\pi \eta ~\omega \left( \frac{\omega}{\omega_c} \right)^{s-1} ~e^{-\omega/\omega_c},
\label{sd}
\end{eqnarray}
where $\eta$ is the coupling strength between the system and the environment, and $\omega_c$ is the frequency cutoff
of the environmental spectra. When $s=1$, $< 1$, and $ > 1$, the corresponding environments are
called as Ohmic, sub-Ohmic, and super-Ohmic in the literature, respectively.
With the spectral density specified, the Green's function $u(t, t_0)$ and $v(t,t+\tau)$ are determined by
Eqs.~(\ref{ide}) and (\ref{vtb}). The non-Markovianity measure can then be calculated from Eq.~(\ref{nonM})
through the two-time correlation functions of Eqs.~(\ref{exact2}) and (\ref{qrtm2}).

\begin{figure}[h]
\includegraphics[width=8.6cm]{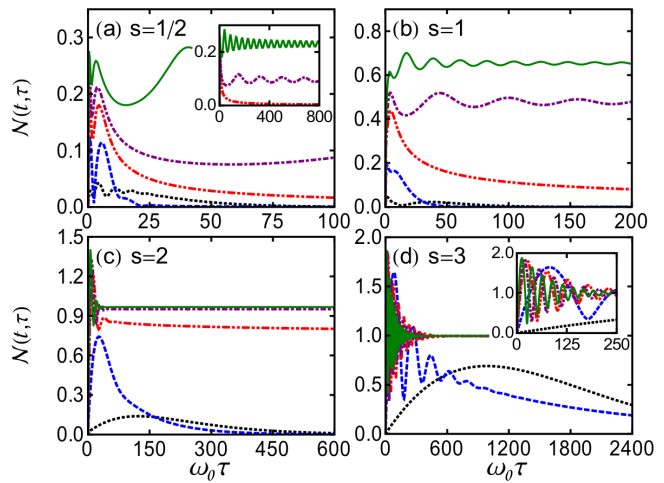}
\caption{\label{fg1} (Color online) The non-Markovianity measure ${\cal N}(t,\tau)$ is plotted as a function of
$\tau$ with $t=0$ at a temperature ($T = 0.5$K) for different system-environment coupling strengths:
$\eta=0.1\eta_c$ (black dotted), $\eta=0.5\eta_c$ (blue dashed), $\eta=1.0\eta_c$ (dashed double dotted red),
$\eta=1.2\eta_c$ (dashed dotted violet), and $\eta=1.5\eta_c$ (solid green).
Four different spectral densities are considered (a) sub-Ohmic ($s=1/2$) (b) Ohmic ($s=1$) (c)
super-Ohmic ($s=2$) and (d) super-Ohmic ($s=3$) for the bosonic reservoir. The other parameters
are taken as $\omega_0=10$ GHz, $\omega_c = 5 \omega_0$, and $n(t_0)=1$.}
\end{figure}

In Fig.~\ref{fg1}, we plot the non-Markovianity measure ${\cal N}(t,\tau)$ as a function of time $\tau$ for
different system-environment coupling strengths $\eta$ at a fixed initial temperature $T = 0.5$K. For simplicity,
we also set $t=0$. The non-Markovianity measure is examined for different spectral densities
with $s=1/2$ (Fig.~\ref{fg1}a for sub-Ohmic), $s=1$ (Fig.~\ref{fg1}b for Ohmic), $s=2$ (Fig.~\ref{fg1}c for super-Ohmic),
and $s=3$ (Fig.~\ref{fg1}d for super-Ohmic). Different curves represent different coupling strengths, namely
$\eta=0.1\eta_c$ (black dotted), $\eta=0.5\eta_c$ (blue dashed), $\eta=1.0\eta_c$ (dashed double dotted red), $\eta=1.2\eta_c$
(dashed dotted violet), and $\eta=1.5\eta_c$ (solid green), where $\eta_c = \omega_0 / {\omega_c \Gamma(s)}$
is the critical coupling strength \cite{general} for the occurrence of the localized mode given in Eq.~(\ref{ut2}).
We observed that for spectral densities with $s \le 2$ and when the coupling strength is below the critical value
(see black-dotted and blue-dashed curves in Figs.~\ref{fg1}a-c for $\eta < \eta_c$), the non-Markovianity measure
increases very fast and approaches a maximum value in a very short time scale, and then decays asymptotically to zero.
This indicate that when the system-environment coupling is weak ($\eta < \eta_c$), the non-Markovianity dynamics
only shows a short-time behavior, as was pointed first by Caldeira and Leggett in quantum Brownian motion \cite{influence2}.
For stronger coupling strengths ($\eta > \eta_c$), the non-Markovianity dynamics is significantly different, and
we find that the non-Markovianity measure tends to approach a saturated value in the long time limit after few slow
oscillations (see dashed-dotted-violet, and solid-green lines in Figs.~\ref{fg1}a-c).
This behavior of non-Markovianity dynamics is contributed from the localized mode given by the
first term in Eq.~(\ref{ut2}). This corresponds to the localized-mode-induced dissipationless dynamics discovered
recently in \cite{general}, as a long-time non-Markovian behavior.

For super-Ohmic case with $s > 2$, the above short-time and long-time behavior of non-Markovianity dynamics
is slightly changed. In this case, for weak coupling $\eta < \eta_c$, the decay rate of the
non-Markovianity measure becomes very slow (see Fig.~\ref{fg1}d) where the non-Markovianity measure
is calculated up to $\omega_0 \tau \sim$ a few thousands. This result is quite different from
the case of $s < 2$ where the non-Markovianity measure already decays to zero at $\omega_0 \tau \sim$
few hundreds or less. This slow decay for $s > 2$ was discussed long time ago in Ref.~\cite{PhysRep},
but in that work \cite{PhysRep}, such a delayed damping was interpreted as an undamped process and they
claimed that the system would never go to its equilibrium for $s > 2$. This interpretation is not fully
correct. As we can see from the black-dotted curve and the blue-dashed curve in Fig.~\ref{fg1}d, it shows
that the non-Markovianity measure damps very very slowly but will eventually decays to zero, so that
the system will approach to equilibrium after a very long time. However, the system will not approach to
equilibrium only in the strong-coupling regime ($\eta > \eta_c$) when the localized mode has significant
contribution \cite{general,thermo1,thermo2,thermo3}. In this case, the non-Markovianity dynamics will show the true
long-time behavior, namely the memory effect will be kept forever. This is given by the dashed-dotted-violet
curve ($\eta = 1.2~\eta_c$) and solid-green curve ($\eta = 1.5~\eta_c$) of Fig.~\ref{fg1}d, where the
non-Markovianity measure reaches a finite value in the long-time limit. To sum up, in Fig.~\ref{fg1},
we show for the first time the existence of both the short time and long time features
of non-Markovianity dynamics for different system-environment coupling strength.

\begin{figure}[h]
\includegraphics[width=8.6cm]{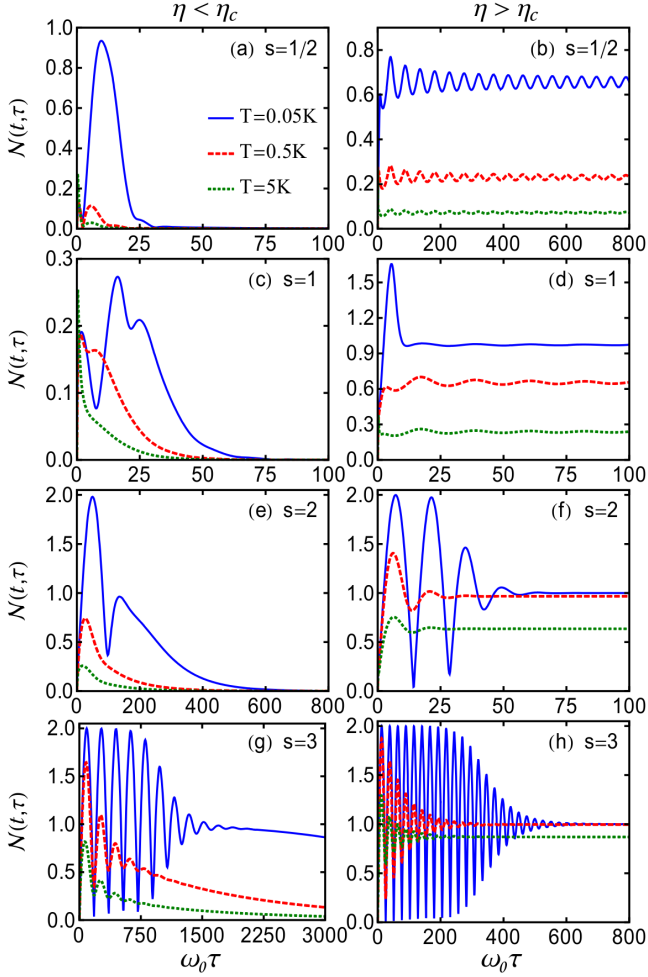}
\caption{\label{fg2} (Color online) Temperature dependence of non-Markovianity measure
is shown as a function of $\tau$ at three different temperatures: $T=0.05$K (solid blue), $T=0.5$K (red dashed),
and $T=5$K (dotted green) with two type of coupling strengths: (a,c,e,g) weak coupling ($\eta=0.5\eta_{c}$) and
(b,d,f,h) strong coupling ($\eta=1.5\eta_{c}$). Four different spectral densities for the bosonic
environment are considered: (a,b) sub-Ohmic ($s=1/2$), (c,d) Ohmic ($s=1$), (e,f) super-Ohmic ($s=2$),
and (g,h) super-Ohmic ($s=3$). The other parameters are taken as $\omega_0=10$ GHz, $\omega_c = 5 \omega_0$,
and $n(t_0)=1$ with a fixed value of $t=0$.}
\end{figure}

In Fig.~\ref{fg2}, we discuss the temperature dependence of the non-Markovianity dynamics using two
different type of coupling strengths: a weak coupling $\eta = 0.5\eta_c$ (see Figs.~\ref{fg2}~a, c, e, and g)
and a strong coupling $\eta = 1.5\eta_c$ (see Figs.~\ref{fg2}~b, d, f, and h). The temperature dependence
of the non-Markovianity originates from the particle number distribution ${\bar n}(\omega,T)$ and non-equilibrium
thermal fluctuation $v(t,t+\tau)$ involved in the two-time correlation functions of Eq.~(\ref{exact2}).
We plot again the non-Markovianity measure for different spectral densities
with $s=1/2$ (Figs.~\ref{fg2}~a,b for sub-Ohmic), $s=1$ (Figs.~\ref{fg2}~c,d for Ohmic),
$s=2$ (Figs.~\ref{fg2}~e,f for super-Ohmic), and $s=3$ (Figs.~\ref{fg2}~g,h for super-Ohmic), where
the temperature dependence of non-Markovianity is given for $T = 0.05$K
(solid blue), $T=0.5$K (red dashed), and $T=5$K (green dotted).
For spectral densities with $s \le 2$ and the coupling strength is weak ($\eta < \eta_c$),
non-Markovianity dynamics still shows a short-time behavior as expected but it takes longer time to decay
to zero at low temperature (see solid-blue curves in Figs.~\ref{fg2}~a, c, and e for $T = 0.05$K).
For a higher temperature, the non-Markovianity decays much faster (see green-dotted curves in
Figs.~\ref{fg2}~a, c, and e for $T=5$K). This indicates that the non-Markovian memory effect is lost
faster when temperature is higher. In stronger coupling regime ($\eta > \eta_c$), non-Markovianity measure
also tends to approach a steady value in the long time limit due to the localized mode contribution but the
magnitude of this stationary values of non-Markovianity become higher for lower temperature of the
environment (see Figs.~\ref{fg2}~b,d). This shows that in both the weak and strong coupling regimes,
thermal fluctuation always reduce the non-Markovian degree for $s < 2$.

For super-Ohmic spectral densities with $s > 2$, the situation is quite different.
For weak coupling $\eta < \eta_c$ and low temperature, the non-Markovianity shows a large oscillation
and then decay extremely slow (see Fig.~\ref{fg2}g for $T = 0.05$K). The damping becomes faster
with a higher temperature of the environment (see the red-dashed and green-dotted curves of Fig.~\ref{fg2}g
for $T = 0.5$K and $T = 5$K). In the strong coupling regime, the non-Markovianity takes much longer
time to reach a steady value which is significantly different from the case for $s < 2$.
In this case, strong oscillations of non-Markovianity is found at low temperature
(solid-blue curve of Fig.~\ref{fg2}h) before approaching its long-time steady value.
The large oscillations persists in the beginning even at higher temperatures (red-dashed and green-dotted lines
of Fig.~\ref{fg2}h). We also notice that the long time stationary values of the non-Markovianity become
less and less sensitive to the temperature, contrary to the case for $s < 2$. To explicitly see this
steady state behavior (see Fig.~\ref{fg2}h), we check the analytic expression of the non-Markovianity
measure, which is greatly simplified for $t=0$ and $\tau\rightarrow\infty$ as
\begin{align}
{\cal N} (0, \tau\rightarrow\infty)\!=\!\Bigg|\frac{u(\tau\rightarrow\infty,0)\sqrt{n(t_0)}}{\sqrt{  n(t_0)
\left| u(\tau\rightarrow\infty,0) \right| ^2 + v(\tau,\tau\rightarrow\infty)}} \Bigg|.
\label{Tempd}
\end{align}
This asymptotic solution shows that the steady value of the non-Markovianity is determined through a
competition between the temperature-dependent thermal fluctuation $v(\tau,\tau\rightarrow\infty)$
and the temperature-independent localized mode contribution $n(t_0)|u(\tau\rightarrow\infty,0)|^2$.
We find that the steady value $u(\tau\rightarrow\infty,0)$ increases as we increase the value of $s$ due to an
enhanced contribution of the localized mode amplitude, while the steady value of thermal fluctuation $v(\tau,\tau\rightarrow\infty)$ decreases with the increase of $s$. This is why the influence of thermal
fluctuation in the steady value of non-Markovianity is less significant for $s > 2$ since the localized
mode contribution dominate over the thermal fluctuation $v(\tau,\tau\rightarrow\infty)$ under this situation.

\begin{figure}[h]
\includegraphics[width=8.6cm]{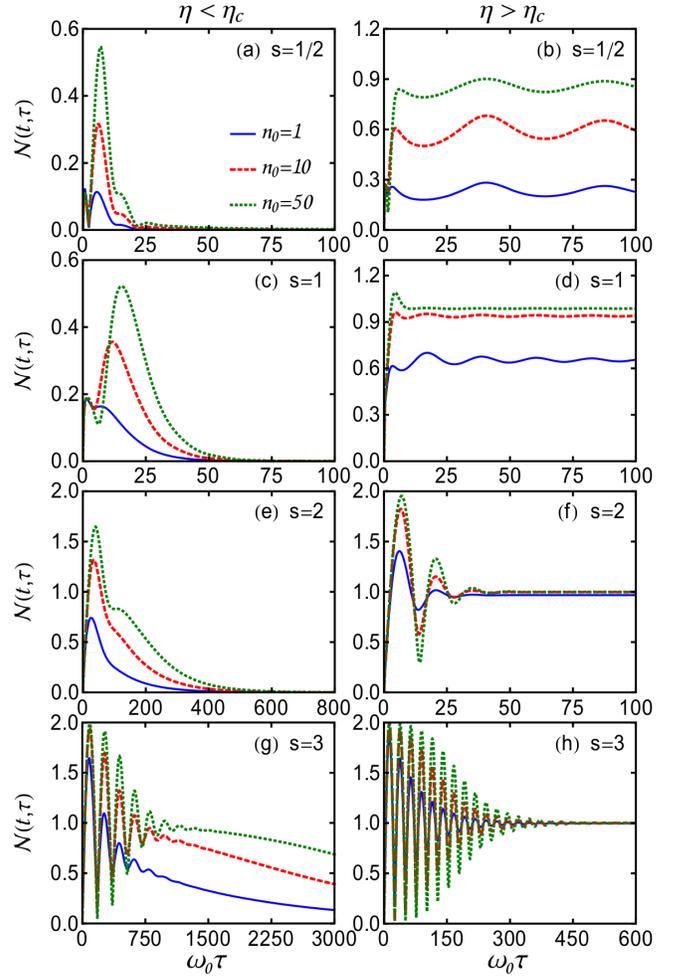}
\caption{\label{fg3} (Color online) Initial state dependence of non-Markovianity measure is
shown as a function of $\tau$ for three different $n(t_0)$: $n(t_0)=1$ (solid blue), $n(t_0)=10$ (red dashed),
and $n(t_0)=50$ (green dotted) with two type of coupling strengths: (a,c,e,g) weak coupling ($\eta=0.5\eta_{c}$) and
(b,d,f,h) strong coupling ($\eta=1.5\eta_{c}$). Four different spectral densities for the bosonic
environment is investigated: (a,b) sub-Ohmic ($s=1/2$), (c,d) Ohmic ($s=1$), and (e,f) super-Ohmic ($s=2$),
and (g,h) super-Ohmic ($s=3$). The other parameters
are taken as $\omega_0=10$ GHz, $\omega_c = 5 \omega_0$, and $T=0.5$K with a fixed value of $t=0$.}
\end{figure}

In Fig.~\ref{fg3}, we show the initial state dependence of non-Markovianity measure.
Here the different initial state is characterized by different initial particle number $n(t_0)$.
We take $n(t_0)=1$ (solid blue), $n(t_0)=10$ (red dashed) and $n(t_0)=50$ (green dotted)
in Fig.~\ref{fg3}, with a fixed temperature ($T=0.5$K) of the environment for weak coupling $\eta = 0.5\eta_c$
(see Figs.~\ref{fg3}~a, c, e, and g) and strong coupling $\eta = 1.5\eta_c$ (see Figs.~\ref{fg3}~b, d, f, and h).
The non-Markovianity dynamics shows an explicit $n(t_0)$-dependence. We find that the magnitude of
non-Markovianity is hiked as the initial particle number $n(t_0)$ is increased in the weak coupling
regime ($\eta < \eta_c$), although the decay time scale of the non-Markovianity looks similar for
different $n(t_0)$ (see Figs.~\ref{fg3}a, c, and e for spectral densities with $s \le 2$). But this
initial state dependence is only a short-time behavior in the weak system-environment coupling regime.
However, for strong coupling ($\eta > \eta_c$) with $s < 2$, we find that the non-Markovianity shows
an explicit long-time dependence of the initial state, due to the existence of the localized modes
in the strong coupling regime (see Figs.~\ref{fg3}b,d for $\eta = 1.5\eta_c$). Thus, the magnitude
of non-Markovianity measure increases in general, as the initial particle number $n(t_0)$ increases.

On the other hand, for super-Ohmic case with $s > 2$ in the weak coupling regime $\eta < \eta_c$, a
very slow decay of the $n(t_0)$-dependent non-Markovianity degree is observed (see Fig.~\ref{fg3}g).
In strong coupling regime, short-time dynamics of non-Markovianity
shows strong oscillations which is $n(t_0)$-dependent (see Fig.~\ref{fg3}h). Whereas the
long-time steady value of non-Markovianity does not significantly depend on $n(t_0)$.
Such a steady state behavior (see Fig.~\ref{fg3}h) is a consequence of the fact that the localized mode contribution
dominate over the thermal fluctuation for $s > 2$, and the non-Markovianity approaches to unity
according to Eq.~(\ref{Tempd}). Physically, the initial state dependence of the non-Markovianity manifests
different non-Markovian memory effects in different time scales. Such non-trivial non-Markovian
behaviors are worthwhile for further experimental tests.


\section{Conclusion}\label{sec:conclusion}

To summarize, we have introduced a new definition of non-Markovianity measure using two-time
correlation functions which play a fundamental role in the exact non-Markovian dynamics of open
quantum systems. Our non-Markovianity measure quantifies the difference between the exact
two-time correlation function and the one obtained in the Markov limit, namely we take
Markov dynamics of two-time correlation function as a reference point for measuring non-Markovianity.
Such a definition of non-Markovianity take into account the proper temperature dependence
given naturally in two-time correlation functions, and can be measured directly in experiments.
We also find that the degree of non-Markovianity is reduced in general due to thermal fluctuations
for hotter environments. Also, our non-Markovianity measure is an explicit function of time.
We discover that in general non-Markovianity shows both the short-time and long-time behavior,
which crucially depends on the system-environment coupling, and thereby the significant contribution
of the localized mode induced by the strong system-environment coupling. The non-Markovianity measure
also depends on the initial state of the open system in different time scales, as a non-trivial
non-Markovian memory effect. Our approach of quantifying
non-Markovianity is applicable to the more general class of open quantum systems
interacting with a general environment through particle tunneling processes.
Further experimental investigations are required to explore these non-trivial dynamics
of the non-Markovianity measure through the measurement of two-time correlation functions.

\begin{acknowledgments}
This work is supported by the National Science Council of Taiwan under Contract No.
NSC-102-2112-M-006-016-MY3 and The National Center for Theoretical Sciences. It is
also supported in part by the Headquarters of University Advancement at the National
Cheng Kung University, which is sponsored by the Ministry of Education, Taiwan, ROC.
\end{acknowledgments}

\appendix

\section{Generalized quantum regression theorem under Born approximation}\label{sec:qrtappendix}

Here we present a generalized quantum regression theorem.
The total density operator of the system plus environment is governed by the quantum
Liouville equation in the Schr\"{o}dinger picture
\begin{eqnarray}
\frac{\partial}{\partial \tau} \rho_{tot}(\tau) = \frac{1}{i\hbar} \left[ H, \rho_{tot}(\tau) \right],
\label{chi1}
\end{eqnarray}
with a formal solution
\begin{eqnarray}
\rho_{tot}(\tau) = e^{-\frac{i}{\hbar}H \tau} \rho_{tot}(0) e^{\frac{i}{\hbar}H \tau}.
\label{chi2}
\end{eqnarray}
Consider an initial state $\rho_{tot}(0)$ of the total system to be uncorrelated and the reservoir
to be in a thermal equilibrium state, that is,
\begin{eqnarray}
\rho_{tot}(0) = \rho(0) \otimes \rho_E(0), ~~~\rho_E(0) = \frac{e^{-\beta H_E}}{\Tr e^{-\beta H_E}},
\label{prod}
\end{eqnarray}
where $H_E = \sum_k \hbar \omega_k b_k^{\dagger} b_k$ and $\beta=1/k_B T$ is the
initial temperature of the reservoir. In this case, if no non-linear dynamics is involved, tracing
over all the environmental degrees of freedom is easily carried out to obtain a time-local master
equation in terms of the reduced density operator
\begin{eqnarray}
\frac{\partial}{\partial \tau} \rho(\tau) = {\cal L}(\tau) \rho(\tau),
\label{Amaster1}
\end{eqnarray}
where $\rho(\tau)=Tr_E[\rho_{tot}(\tau)]$. One can then calculate the mean values of a complete set of system
operators ${\hat O}_i$,  $i = 1, 2, ...$, using Eq.~(\ref{Amaster1}). The expectation value of an operator
${\hat O}_i$ obey the relation
\begin{eqnarray}
\hskip -0.6cm
\frac{\partial}{\partial \tau}
\langle {\hat O}_i(\tau) \rangle = \langle {\hat O}_i(0) {\dot \rho}(\tau) \rangle
= \langle {\hat O}_i(0) {\cal L}(\tau) \rho(\tau) \rangle.
\label{Amean1}
\end{eqnarray}
For each ${\hat O}_i$, one can define
\begin{eqnarray}
\langle {\hat O}_i(0) {\cal L}(\tau) \rho(\tau) \rangle = \sum_j M_{ij}(\tau) \langle {\hat O}_j(\tau) \rangle,
\label{Amean2}
\end{eqnarray}
Then Eq.~(\ref{Amean1}) becomes
\begin{eqnarray}
\frac{\partial}{\partial \tau} \langle {\hat O}_i(\tau) \rangle = \sum_j M_{ij}(\tau) \langle {\hat O}_j(\tau) \rangle.
\label{Amean3}
\end{eqnarray}

On the other hand, given two system operators ${\hat O}_1$ and ${\hat O}_2$, their two-time correlation function
is defined more conveniently in the Heisenberg picture,
\begin{eqnarray}
\hskip -0.5cm
\langle {\hat O}^{H}_1(t_1) {\hat O}^{H}_2(t_2)  \rangle = \Tr_{S \oplus E}
\left [ \rho^{H}_{tot} {\hat O}^{H}_1(t_1) {\hat O}^{H}_2(t_2)  \right].
\label{step1}
\end{eqnarray}
Then going back to the Schr\"{o}dinger picture through the transformations
\begin{eqnarray}
{\hat O}^{H}_1(t_1) &=& e^{\frac{i H t_1}{\hbar}} {\hat O}_1 e^{-\frac{i H t_1}{\hbar}}, \\
{\hat O}^{H}_2(t_2) &=& e^{\frac{i H t_2}{\hbar}} {\hat O}_2 e^{-\frac{i H t_2}{\hbar}}, \\
\rho^{H}_{tot} &=& e^{\frac{i H t}{\hbar}} \rho_{tot}(t) e^{-\frac{i H t}{\hbar}},
\end{eqnarray}
also taking $t_1=t$ and $t_2=t+\tau$, and using the cyclic property of trace, we obtain
\begin{eqnarray}
\nonumber
\hskip -0.5cm
\langle {\hat O}^{H}_1(t) {\hat O}^{H}_2(t+\tau)  \rangle &=& \Tr_{S \oplus E} \left [ {\hat O}_2~
e^{-\frac{i H \tau}{\hbar}} \rho_{tot}(t) {\hat O}_1  e^{\frac{i H \tau}{\hbar}} \right] \\
&=& \Tr_{S \oplus E} \left [ {\hat O}_2 ~\rho_{tot,{\hat O}_1}(t,\tau) \right],
\label{step2}
\end{eqnarray}
where we have used the fact that the state of the total system $\rho_{tot}(t)$ in the Schr\"{o}dinger picture
acted by the operator ${\hat O}_1$ becomes a new state denoted by
$\rho_{tot,{\hat O}_1}(t,0)\equiv \rho_{tot}(t){\hat O}_1$ such that
\begin{eqnarray}
\hskip -0.7cm
\rho_{tot,{\hat O}_1}(t,\tau) = e^{-\frac{i H \tau}{\hbar}} \rho_{tot,{\hat O}_1}(t,0) e^{\frac{i H \tau}{\hbar}}.
\label{step3}
\end{eqnarray}
Now, if $\rho_{tot}(t)$ is factorized as
\begin{eqnarray}
\rho_{tot}(t)=\rho(t) \otimes \rho_E(0),
\label{Afacto}
\end{eqnarray}
then
\begin{eqnarray}
\hskip -0.6cm
\rho_{tot,{\hat O}_1}(t,0) = \rho_{{\hat O}_1}(t,0) \otimes \rho_E(0),
\label{Aasmp}
\end{eqnarray}
where $\rho_{{\hat O}_1}(t,0)\equiv \rho(t){\hat O}_1$ is the state of the system acted
by ${\hat O}_1$ at time $t$. Thus, Eq.~(\ref{step2}) can be reduced to
\begin{eqnarray}
\nonumber
\langle {\hat O}^{H}_1(t) {\hat O}^{H}_2(t+\tau)  \rangle
&=& \Tr_{S} \left [ {\hat O}_2 \Tr_{E}\{\rho_{tot,{\hat O}_1}(t,\tau)\} \right] \\
&=& \Tr_{S} \left [ {\hat O}_2 ~\rho_{{\hat O}_1}(t,\tau) \right],
\label{step99}
\end{eqnarray}
with $\rho_{{\hat O}_1}(t,\tau)=\Tr_{E}\{\rho_{tot,{\hat O}_1}(t,\tau)\}$.

As one can see, Eq.~(\ref{Aasmp}) for $\rho_{tot,{\hat O}_1}(t,0)$ obeys the same initially
factorized condition of Eq.~(\ref{prod}) for $\rho_{tot}(0)$, and the formal solution (\ref{step3})
for $\rho_{tot,{\hat O}_1}(t,\tau)$ has the same form of Eq.~(\ref{chi2}) for $\rho_{tot}(\tau)$.
Thus, the master equation for $\rho_{{\hat O}_1}(t,\tau)$ with respect to $\tau$ must be
the same as that for $\rho(\tau)$, namely
\begin{eqnarray}
\frac{\partial}{\partial \tau} \rho_{{\hat O}_1}(t,\tau)
= {\cal L}(\tau) \rho_{{\hat O}_1}(t,\tau).
\label{master9}
\end{eqnarray}
This gives
\begin{eqnarray}
\frac{\partial}{\partial \tau} \langle {\hat O}^{H}_1(t) {\hat O}^{H}_2(t+\tau) \rangle =
\langle {\hat O}_2 {\cal L}(\tau) \rho_{{\hat O}_1}(t,\tau) \rangle.
\label{step5}
\end{eqnarray}
For ${\hat O}_2$, we choose a complete set of system operators ${\hat O}_i$ so that for each ${\hat O}_i$
\begin{eqnarray}
\hskip -0.7cm
\langle {\hat O}_i(0) {\cal L}(\tau) \rho_{{\hat O}_1}(t,\tau) \rangle = \sum_j M_{ij}(\tau) \langle {\hat O}_j
\rho_{{\hat O}_1}(t,\tau) \rangle.
\label{step6}
\end{eqnarray}
Thus, we reach
\begin{eqnarray}
\nonumber
\hskip -0.5cm
\frac{\partial}{\partial \tau} \langle {\hat O}^{H}_1(t) {\hat O}^{H}_i(t+\tau)  \rangle =
\sum_j M_{ij}(\tau) \langle {\hat O}^{H}_1(t) {\hat O}^{H}_j(t+\tau) \rangle. \\
\label{step7}
\end{eqnarray}
This is the generalized quantum regression theorem namely, the form of the evolution equation (\ref{Amean3})
for single-time expectation values is the same as that for the two-time correlation functions (\ref{step7}).
Notice that Eq.~(\ref{Aasmp}) is the well-known Born approximation in the derivation of the perturbative
master equation up to the second order for system-environment coupling. Thus, the above generalized quantum
regression theorem is only valid under Born approximation.

\section{Calculating $\langle a^{\dagger}(t) a(t+\tau) \rangle_{qrt}$}\label{sec:qrtexac}

Using the exact master equation (\ref{master2}), one can obtain the evolution matrix $M_{ij}(\tau)$ of
Eq.~(\ref{Amean3}) by calculating the single-time expectation values of a complete set of system operators
($a$, $a^{\dagger}$, $n$ and $\mathbb{1}$), where $n=\langle a^{\dagger} a \rangle$:
\begin{equation}
{\bf M}(\tau) =
\left( \begin{array}{cccc}
\alpha(\tau) & 0 & 0  & 0 \\
0 & \alpha^{\ast}(\tau) & 0 & 0 \\
0 & 0 & -2\gamma(\tau) & \widetilde{\gamma}(\tau) \\
0 & 0 & 0 & 0
\end{array} \right),
\label{matrix}
\end{equation}
where $\alpha(\tau) = -\left( \gamma(\tau) + i \omega_0^{\prime}(\tau) \right)$. The evolution matrix
elements of ${\bf M}(\tau)$ are now time-dependent, which is different from the Markov evolution where
the matrix elements are time-independent, see Eq.~(\ref{matrix2}). The time-dependent coefficients in
the exact master equation (\ref{master2}) are explicitly given by
\begin{eqnarray}
\label{cft1}
\omega_0^{\prime}(\tau) &=& -Im\left[\frac{{\dot u}(\tau,t_0)}{u(\tau,t_0)}\right], \\
\label{cft2}
\gamma(\tau) &=& -Re\left[ \frac{{\dot u}(\tau,t_0)}{u(\tau,t_0)} \right], \\
\label{cft3}
\widetilde{\gamma}(\tau) &=& {\dot v}(\tau,\tau) + 2 v(\tau,\tau) \gamma(\tau).
\end{eqnarray}
Now, naively applying generalized quantum regression theorem as done in Ref.~\cite{nvtyVacchini2014}
(this is indeed incorrect because the exact master equation does not involve the Born approximation),
the evolution equation for the two-time correlation function
$\langle a^{\dagger}(t) a(t+\tau) \rangle$ is given by
\begin{eqnarray}
\nonumber
&&\frac{d}{d\tau} \langle a^{\dagger}(t) a(t+\tau) \rangle_{qrt} \\
\nonumber
&&{} = -\left( \gamma(\tau) + i \omega_0^{\prime}(\tau) \right) \langle a^{\dagger}(t) a(t+\tau) \rangle_{qrt} \\
&&{} = \frac{{\dot u}(\tau,t_0)}{u(\tau,t_0)} \langle a^{\dagger}(t) a(t+\tau) \rangle_{qrt}.
\label{Bqrte1}
\end{eqnarray}
The exact solution of this equation is
\begin{eqnarray}
\langle a^{\dagger}(t) a(t+\tau) \rangle_{qrt} = n(t) u(\tau,t_0),
\label{Bsoln}
\end{eqnarray}
where $n(t)=\langle a^{\dagger}(t) a(t)\rangle$ which is given by \cite{bosonic}
\begin{equation}
n(t) = v(t,t) + u^{\ast}(t,t_0) n(t_0) u(t,t_0).
\label{pn}
\end{equation}
Thus, the two-time correlation function naively using the generalized quantum regression theorem
is given by
\begin{eqnarray}
\nonumber
\langle a^{\dagger}(t) a(t+\tau) \rangle_{r} &=& u^{\ast}(t,t_0) n(t_0) u(t,t_0) u(\tau,t_0) \\
&&{} +  v(t,t) u(\tau,t_0).
\label{Bqrte3}
\end{eqnarray}
This solution is not the exact two-time correlation function given by Eq.~(\ref{exact2}) nor
the result of two-time correlation function in the Markov limit, {\it i.e.} Eq.~(\ref{qrtm2}).

\end{document}